\documentclass{evn2004}
\usepackage{txfonts}
\usepackage{graphicx}
\begin{document}

\title{The simultaneous VLA observations of Sgr A* from 90 to 0.7 cm}

\author{T. An \inst{1} \and J.-H. Zhao\inst{2} \and X.-Y. Hong\inst{1} \and Z.-Q. Shen\inst{1}
 \and W.M. Goss\inst{3} \and
S.Roy \inst{4} \and A.P. Rao\inst{4}}

\institute{Shanghai Astronomical Observatory, Chinese Academy of Sciences, Shanghai
200030, China \and Harvard-Smithsonian CfA, 60 Garden St, MS 78, Cambridge, MA 02138, USA \and
NRAO-AOC, P. O. Box 0, Socorro, NM 87801, USA \and National Center for Radio Astrophysics
(TIFR), Pune University Campus, Post Bag No. 3, Ganeshkhind, Pune 411 007, India}

\abstract{ We present a spectrum of Sgr A* observed simultaneously on June 17,
2003 at wavelengths from 90 to 0.7 cm with the VLA. In the spectrum, we also
include the measurements of Sgr~A* observed on the same day with the GMRT at 49
cm, the SMA at 0.89 mm and the Keck II at 3.8 $\mu$m. The spectrum at the
centimeter wavelengths suggests the presence of a break wavelength at
$\lambda_{\rm b}\sim$3.8 cm (8 GHz). The spectral index is $\alpha=0.43\pm0.03$
(${\rm S\propto\nu^\alpha}$) at 3.8 cm and shorter wavelengths. The spectrum
between $\lambda =$3.8 cm and $\lambda =$49 cm can be described by a power law
with spectral index of $\alpha=0.10\pm0.03$. We detected Sgr A* with
$0.22\pm0.06$ Jy at 90 cm, suggesting a sharp decrease in flux density at the
wavelengths longer than 49 cm. The best fit to the spectrum at the wavelengths
longer than $\lambda_{\rm b}$ appears to be consistent with free-free
absorption by a screen of ionized gas with a turnover wavelength at $\nu
(\tau_{\rm ff}=1)\sim 100$ cm (300 MHz). This turnover wavelength appears to be
three times longer than that of 30 cm (1 GHz) as suggested by Davies et al.
(\cite{Davies}) based on the observations in 1994 and 1995. Our analysis
suggests that stellar winds from the massive stars near Sgr A* could modulate
the flux density at the wavelengths longer than  30 cm (or frequencies below 1
GHz).}

\maketitle

\section{Introduction} \label{intro}

The compact radio source Saggittarius  A* (Sgr A*) at the Galactic center (GC)
is widely believed to be associated with the supermassive black hole (SMBH)
with a mass of $M=4\times10^6 M_{\odot}$ (Sch\"{o}del et al. \cite{Schodel};
Ghez et al. \cite{Ghez03}). Sgr A* is a prototypical case of  low luminosity
galactic nuclei. Theoretical models suggest that a model with low efficiency
radiative accretion flow might explain the dim nature of Sgr A* based on the
observations at the wavelengths of radio sub-millimeter, IR and X-rays. In a
recent paper, Loeb (\cite{Loeb}) proposed that the stellar wind streams from
the young, massive stars may also play an important role in fueling Sgr A* in
order to explain the variability in flux densities observed at
millimeters/submillimeters. However, the ionized gas of the stellar winds may
also attenuate the flux density at longer radio wavelengths. On the other hand,
Davies et al. (\cite{Davies}, hereafter DWB) observed Sgr A* at 0.408, 0.96 and
1.66 GHz and only detected at the two higher frequencies. They suggested a
low-frequency cutoff at about 1 GHz and suggested that the non-detection at 408
MHz was owing to the free-free absorption. However, detections of significant
flux density from Sgr A* at 620 MHz with the GMRT (Roy \& Rao \cite{Roy1}) and
at 330 MHz with the VLA (Nord et al. \cite{Nord}) have raised questions of the
early observations and interpretations by Davies et al. (\cite{Davies}).

\section{Observations}
We carried out simultaneous observations of Sgr A* with the VLA in its A
configuration on June 17, 2003, covering a wide frequency range of 90, 20, 6,
3.6, 2, 1.3 and 0.7 cm. The observations at 90 and 20 cm were done in spectral
line mode in order to reject RFIs and to minimize the bandwidth smearing
effect. Observations at other bands were performed in the VLA standard
continuum mode. In this paper, we present the preliminary results from our
observations and analysis. The detailed results will be given in An et al.
(\cite{An}).

\section{Results and discussion}

\subsection{Detection of Sgr A* at 90 cm}\label{Dectection}

We carried out several iterations of phase-only self-calibration to the 90 cm
data in order to improve the dynamic range of the image (please see electronic
version of the image at 90 cm). The final image is restored with a beam of
$10.9''\times6.8''$ (PA$=-10\degr$). The off-source \emph{r.m.s.} noise of
$\sim12$ mJy/b in the image is estimated. Sgr A East dominates the total flux
density at this wavelength.
A bump in flux density is observed at the expected location of Sgr
A*. The region surrounding  Sgr A*, except for the southern extended emission,
is significantly absorbed by the ionized gas in Sgr A West. We carried out
Gaussian fitting with the intensity slices along the expected major and minor
axes to estimate the flux density at 90 cm. The fitted peak brightness is
65$\pm$16 mJy/b after subtracting the background confusion, and the deconvolved
source size is $14.4''\times10.7''$. A total flux density of $220\pm60$ mJy at
90 cm is derived from the model fitting.

\subsection{Simultaneous Spectrum at Radio Wavelengths}

\begin{figure}
\includegraphics[width=0.5\textwidth]{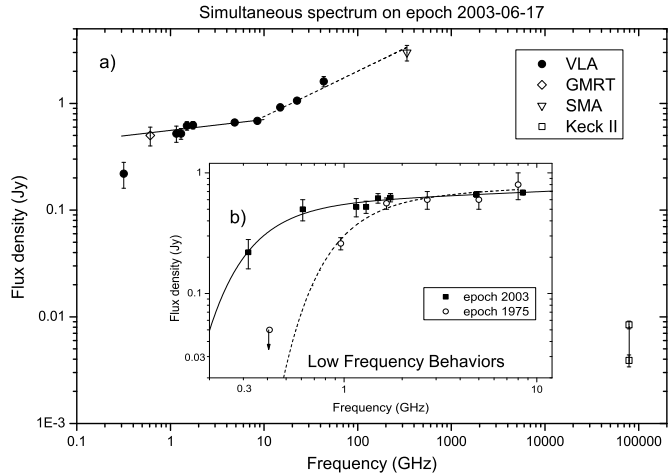}\vspace{-3mm}
\caption{\textbf{a)} : the observed spectrum of Sgr A* from 90 cm to 3.8$\mu$m.
\emph{solid circle}: VLA from 90 to 0.7 cm (present paper); \emph{diamond}:
GMRT at 610 MHz (Roy \& Rao \cite{Roy2}); \emph{triangle}: SMA at 335 GHz (Zhao
et al. \cite{Zhao04}) and \emph{square}: Keck II at 3.8$\mu$m (Ghez et al.
\cite{Ghez04}).  \textbf{b)} : comparison of low frequency free-free fitting
between epoch 2003 and epoch 1975. \emph{solid square} : epoch 2003; \emph{open
circle} : epoch 1975. }\label{fig:spec}
\end{figure}

We also measured the flux densities of Sgr A* at other shorter wavelengths in
the IMAGE domain. We use a point-source model (for wavelengths at 6 cm and
shorter) or an elliptical Gaussian model (for wavelengths $\sim$20 cm) to fit
the data. The spectrum shape around 1 GHz is critical in  examination of  the
low frequency turnover. Thus, we measured the flux density at four discrete
frequencies (1.155, 1.300, 1.500 and 1.740 GHz) in the 20 cm band. We also
measured the flux density on each band in the UV domain. The difference between
the flux densities derived from the two procedures was taken as the uncertainty
in measurements ($\Delta S_M$). Taking into account of calibration uncertainty
($\Delta S_C$), the inferred 1 $\sigma$ errors   are the quadrature addition of
two terms. The uncertainties in a fraction of the flux densities are 27\%,
17\%, 11\%, 9.1\%, 7.6\%, 5.4\%, 4.5\%, 7.0\%, 5.6\% and 11\% at the
wavelengths between 90, 26, 23, 20, 17, 6, 3.6, 2, 1.3 and 0.7 cm,
respectively.

Figure \ref{fig:spec}a shows the a quasi-simultaneous (within a day) spectrum
of Sgr A* from 90 cm to 3.8 $\mu$m covering 5 orders of magnitude in the
frequency range. A significant break in the cm-band spectrum is seen around
$\lambda_b\sim3.8$ cm (8 GHz). For the wavelengths shorter than $\lambda_b$ the
spectrum can be described by a power-law spectrum:
$S_\nu\propto\nu^{0.43\pm0.03}$. The spectrum appears to peak at a
millimeter/submillimeter wavelength and the spectral index determined from the
flux densities between  0.87 mm and 3.8 $\nu$m is $\alpha_{0.87mm/3.8\mu
m}=-1.13^{+0.06}_{-0.08}$.

The dashed line represents the fitting between 8 and 335 GHz. The spectrum
between 49 cm (620 MHz) to 3.8 cm (8 GHz) is rather flat with a spectral index
$\alpha=0.10\pm0.03$. At 318 MHz ($\sim90$ cm), the spectrum shows a factor of
$\sim$2 decrease in flux density in comparison with the value observed at 49
cm.

\subsection{Low frequency behaviors}

In Figure \ref{fig:spec}b, we compare the low-frequency (from 318 MHz to 8.5
GHz) spectrum measured on the epoch of 1975 (DWB; \emph{open circle}) and on
the epoch of 2003 (\emph{solid square}). DWB showed an exponential cut-off in
flux densities of Sgr A* below 1 GHz and suggested that the cut-off is owing to
the free-free absorption by the ionized gas in Sgr A West (the dashed line
represents the free-free fitting). However the free-free absorption model
appears to confront with the GMRT observation at 49 cm (Roy \& Rao \cite{Roy2})
and our VLA observation at 90 cm in 2003 shown in Fig. \ref{fig:spec}b. The
flux densities of Sgr A* determined from both the GMRT observations at 49 cm
and the VLA observations at 90 cm are significantly greater than the values
expected from DWB's free-free absorption model. However the  measurements from
the nearly simultaneous observations using the GMRT and the VLA do show a
significant decrease in flux density at 90 cm as is compared to that at 49 cm.

It is unlikely that the flux density variations at the lower frequencies are
intrinsic to Sgr A* since the source appears to be quiet at the shorter
centimeters based on the monitoring observations (Zhao et al. \cite{Zhao01}).
If we believe DWB's measurements and free-free absorption model is correct,
then the variations in the lower frequencies would suggest that the column
density of the free-free absorbing screen is changed over the past 30 years. We
can fit DWB's free-free absorption model to the data observed in 2003 in order
to assess the change in the column density of the ionized in the front of Sgr
A*. The solid line in Figure \ref{fig:spec}b shows the free-free fit to the
2003 data  and the dashed line is the fit to the 1975 data. The discrepancy in
free-free fitting between epoch 1975 and 2003 suggests that the critical
cut-off frequency ($\nu(\tau_{ff}=1)$) in free-free absorption must shift from
$\sim$1 GHz to $\sim$300 MHz in past 30 years. The inferred shift
$\Delta\nu(\tau_{ff}=1)$ suggests that the free-free optical depth at a given
frequency has decreased by a factor of 9. Assuming the electron temperature in
the ionized gas did not change in the past 30 years, the inferred decrease in
optical depth could be due to a decrease of electron column density  in the
front of Sgr A*. Such a large-scale variation in the column density of
electrons  in a time-scale of 30 years was likely taken place in a compact
region near Sgr A*. The stellar winds from the orbiting massive stars around
Sgr A* within the central 1\arcsec could modulate the flux density at low radio
frequencies from Sgr A* on a timescale of 10 years (Loeb \cite{Loeb}).

As a synchrotron source, the low frequency behavior of Sgr A* could also be
explained by synchrotron self-absorption model (SSA), which is sensitive to the
critical cut-off frequency. However, the SSA model requires the strength of the
magnetic field much higher than the value that is common accepted.

The Razin suppression effect suggests a cut-off frequency at $10^{1-2}$ kHz
(Davies et al. \cite{Davies}) which is far below the observed value of
$\sim300$MHz.


\end{document}